\documentstyle[12pt]{article}
\newcommand \half {{\textstyle {1\over 2}}}

\begin{document}
\hsize=6.5truein
\hoffset=-.8truein

\begin{titlepage}
\null\vspace{-62pt}

\pagestyle{empty}
\begin{center}
\rightline{CCNY-HEP-99/4}
\rightline{hep-ph/9905569}

\vspace{1.0truein}
{\Large\bf Plasmon interactions in the quark-gluon plasma }

\vspace{1in}
D. METAXAS and V.P. NAIR\\
\vskip .4in
{\it Physics Department\\
City College of the City University of New York\\
New York, NY  10031\\
metaxas@ajanta.sci.ccny.cuny.edu, vpn@ajanta.sci.ccny.cuny.edu}\\

\vspace{0.5in}

\vspace{.5in}
\centerline{\bf Abstract}

\baselineskip 18pt
\end{center}

Yang-Mills theory at finite temperature is rewritten as a theory of plasmons
which provides a Hamiltonian framework for perturbation theory with resummation of hard
thermal loops.

\end{titlepage}

\newpage
\pagestyle{plain}
\setcounter{page}{2}
\newpage

\section{Introduction}

It is by now a well-recognized fact that the so-called hard thermal loops (HTL's) play
a very important role in Yang-Mills theories at high temperatures \cite{BP1}.
Hard thermal loops and the corresponding effective action have been studied rather
extensively over the last few years \cite{Taylor, VPN1, blaizot, manuel1}.
Hard thermal loops describe Debye screening, Landau damping, etc., and a resummation of
perturbation theory including their effects is essential to avoiding a
class of infrared divergences due to interactions of the electrostatic type. While this is
clear conceptually, the actual implementation of a resummed perturbation theory has not been
easy \cite{tft98, braat}. Further, even in a
resummed perturbation theory, there are still infrared problems which can be cured 
only by
incorporating magnetic screening effects as well \cite{GPY}. Such effects are expected to be
of order $g^2 T$ where $g$ is the coupling constant and $T$ is the temperature.
In order to analyze magnetic screening effects, one must first incorporate
hard thermal loop effects (which are of order $gT$ ) and obtain an effective theory 
valid at lower scales, of order $g^2T$. There have been many related approaches 
to this question in recent literature, based on kinetic equations, thermal Feynman
diagrams, etc. \cite{iancu, bodeker, manuel2}.
In this connection, it is worth emphasizing that magnetic screening involves the wave
properties of the gauge fields. The magnetic mass may be considered as the (dynamically
generated) mass gap
of the effective three-dimensional theory of the zero Matsubara frequency modes; 
in terms of counting of powers of $\hbar$ this would  seem to be a classical effect. 
However,
the relevant classical theory is a wave theory, not a particle theory. In particular 
it is easy
to see that, in the low energy limit where the zero Matsubara frequency approximation 
is valid,
the single-particle wavefunctions have significant overlap, similar to 
what happens in Bose-Einstein
condensation. Classical particle descriptions or corresponding kinetic equations will not
suffice to generate a magnetic mass. One would thus like to have a way of incorporating the
HTL-effects within a formalism which can systematically treat dynamical correlation
effects
as well as thermal corrections and preferably quantum corrections as well. The natural
candidate for this would be an action formalism where one can add and subtract electric and
magnetic mass terms and develop a resummed perturbation theory. While this is systematic,
despite calculational complexity, the effective action for hard thermal loops involves
nonlocality in time and some conceptual issues are better addressed in  a Hamiltonian
approach. Since the theory at finite temperature lacks manifest covariance anyway, there seems
to be no serious drawback to a Hamiltonian analysis. In this paper we set up the basic
framework of a resummed Hamiltonian analysis. 

The perturbative eigenmodes of the plasma of gluons at finite temperature are the plasmons.
In addition to the two transverse polarizations, there is also a longitudinal
polarization, which is physical mode at finite temperature, obeying the Gauss law.
Our approach is to rewrite the theory as a theory of such plasmons. 
We work out interactions of the plasmons to the quadratic order in the coupling constant.
The Coulomb interaction between plasmons shows the screened behaviour, as expected.
Perturbative calculations with this Hamiltonian will be a HTL-resummed perturbation theory.
This sets up the basic framework. The next step is to use this to calculate 
corrections to HTL-effects, some of which are under way.

\section{Hamiltonian analysis}

As mentioned in the introduction,
our approach will be to simplify the Hamiltonian analysis.
The action for a non-abelian gauge theory, with the
HTL-effective action added, is given by \cite{VPN1}
\begin{eqnarray}
S=-\frac{1}{4} \int F^2 &+& \kappa \int d\Omega
         \Bigl[ d^2 x^T S_{WZW}(G) + 
          \frac{1}{\pi}\int d^4 x \, {\rm Tr} (G^{-1} \partial_{-}G\, A_{+}
            \nonumber \\
 & &  - A_{-}\partial_{+} G G^{-1} \, + A_{+} G^{-1} A_{-} G \, -
                        A_{+}A_{-})\Bigr] 
\label{WZW}
\end{eqnarray}
where $S_{WZW}(G)$ is the standard Wess-Zumino-Witten (WZW) action for $G$.
$G(x, Q)$ is a unitary matrix field depending on
$x^{\mu}$ and on the unit vector $\vec{Q}$, $\vec{Q}\cdot {\vec Q}=Q^2 = 1$,
and obeying
$G^{\dagger}(x, \vec{Q})=G(x, -\vec{Q})$.
Also in (\ref{WZW}),  
$\partial_{\pm}=
\half (\partial_{0}\pm \vec{Q}\cdot\vec{\partial} )$ and
$A_{\pm}=
\half ( A_{0}\pm \vec{Q}\cdot\vec{A})$.
The $G$-dependent term of the action includes
integration over the angles of $\vec{Q}$.
The parameter $\kappa$ is given in the lowest order
calculation by
$(N + \half N_{F}) T^2 /6$,
where $N$ is the number of colors, $N_{F}$
is the number of quark flavors and $T$ is the temperature.
For our purpose,  $\kappa$ can be considered as a
free parameter.

The Hamiltonian analysis of (\ref{WZW}) has been
given elsewhere \cite{VPN2}.
With the gauge choice of $A_{0}=0$, the Hamiltonian
is given by
\begin{equation}
H=\int d^3 x \,\frac{E^2 + B^2}{2} \,+\,
  \frac{2\pi}{\kappa}\int d^3 x \,d\Omega ~(J_+^2 + J_-^2)
\label{Ham}
\end{equation}
where $E_i^a = \partial_0 A_i^a $
and
$B_i^a = \epsilon_{ijk}(\partial_j A_k^a + \frac{1}{2}
                              f^{abc} A_b^j A_c^k ) $
are the usual nonabelian electric 
and magnetic fields.
The currents $J_{\pm}$ are given by
\begin{eqnarray}
J_+ &=& \frac{\kappa}{4\pi} D_+ G \, G^{-1} = -i t^a J_+^a 
\nonumber \\
J_- &=& -\frac{\kappa}{4\pi} G^{-1}D_- G = -i t^a J_-^a
\label{def}
\end{eqnarray}
and are related by $J_+(x, -\vec{Q})=J_-(x, \vec{Q})$.

The equal-time commutator  algebra which supplements 
the Hamiltonian (\ref{Ham}) is given by
\begin{eqnarray}
{[E_i^a (\vec{x}), A_j^b (\vec{y})]} &=& -i \delta^{ab}\delta_{ij}
                                          \delta(\vec{x}-\vec{y})  
                                             \nonumber \\
{[E_i^a (\vec{x}), J_{+}^{b} (\vec{y})]} &=&  i \frac{\kappa}{4\pi}
                          Q_i \delta^{ab} \delta(\vec{x}-\vec{y})
                               \nonumber \\
{[J_+^a (\vec{x}, \vec{Q}), J_+^b (\vec{y}, \vec{Q'})]} &=& 
- i \frac{\kappa}{4\pi}
(Q\cdot\nabla_x \delta^{ab} - f^{abc} A^c (x))
\delta (\vec{x}-\vec{y})\delta(Q,Q')
\nonumber \\
& &
+i f^{abc} J_+^c (\vec{x}, \vec{Q})
\delta (\vec{x}-\vec{y}) \delta(Q,Q')
\label{opal}
\end{eqnarray}
In addition to the  Hamiltonian (\ref{Ham}) and the
operator algebra (\ref{opal}), we must
also impose the Gauss law as a constraint
selecting the physical states  $ |\psi\rangle $, i.e.,
we must require that ${\cal{G}}^a  |\psi\rangle = 0  $, where
\begin{equation}
{\cal{G}}^a = (\vec{D}\cdot E)^a +
            \int d\Omega ~(J_+^a + J_-^a)
\label{Gauss}
\end{equation}
Equations (\ref{Ham}, \ref{opal}, \ref{Gauss}) 
define the theory.
The rest of this paper will be devoted to the
analysis of these equations.

The currents $J_{\pm}^a$ obey the generalized
Kac-Moody algebra of (\ref{opal}). The first step
in our approach will be to introduce
a canonical set of variables
$\phi^a (x, \vec{Q})$, $\Pi^b (x, \vec{Q})$
that depend on $\vec{Q}$ as well as $\vec{x}$,
which obey the canonical algebra
\begin{equation}
{[\phi^a (\vec{x}, \vec{Q}),  \Pi^b (\vec{y}, \vec{Q'})]} 
= -i \delta^{ab}
    \delta(\vec{x}-\vec{y})
      \delta(\vec{Q}, \vec{Q'})
\label{canon}
\end{equation}
with all other commutators
equal to zero.

We are interested in solving the theory to a certain
order in the coupling constant or power of the
structure constants $f^{abc}$, and so we can
solve (\ref{opal}) as a series in $f^{abc}$.
The solution for $J_+^a$ is given by
\begin{eqnarray}
J_+^a =& &  \Pi^a - \frac{\kappa}{8\pi} Q\cdot \nabla \phi^a
               - \frac{\kappa}{4\pi} Q\cdot A^a
                  \nonumber \\
        &-& \frac{\kappa}{48\pi} f^{abc} \phi^b Q\cdot \nabla \phi^c
        + \frac{1}{2} f^{abc} \phi^b \Pi^c 
                 \nonumber \\
       & +& \frac{1}{24}f^{abc}f^{brs}(\Pi^r \phi^s \phi^c
                                     + \phi^c \Pi^r \phi^s)+\cdots
\label{current}
\end{eqnarray} 
It is easily verified that this solves
the algebra (\ref{opal}) to the quadratic order in $f^{abc}$.
This also agrees with the expression (\ref{def}) for $J_+^a$
in a suitable parametrization of $G(x, \vec{Q})$ 
in terms of $\phi^a (x, \vec{Q})$.
$J_-^a$ is given by $\vec{Q}\rightarrow -\vec{Q}$
in (\ref{current}).
Since the unitary matrix  $G(x, \vec{Q})$
obeys $G^{\dagger}(x, \vec{Q}) = G(x, -\vec{Q})$
we must have $\phi(x, -\vec{Q})= -\phi(x, \vec{Q})$
and $\Pi (x, -\vec{Q})= -\Pi (x, \vec{Q})$.
(Strictly speaking, this condition has to be imposed as a weak condition; however, as
the comments in the concluding section make clear, this will not affect the reduction of the
Hamiltonian which follows.)
By direct substitution from (\ref{current}) and keeping in mind the
above mentioned property, we find the Hamiltonian to quadratic order
in $f^{abc}$ as 
\begin{eqnarray}
{\cal{H}}&& = \int d^3 x \Biggl[ \frac{E^2 + B^2}{2} 
            + \frac{4\pi}{\kappa}\int_{\Omega}
                (\Pi - \frac{\kappa}{4\pi}Q\cdot A)^2
 +{\kappa\over 16\pi }\int_{\Omega}(Q\cdot\nabla \phi)^2
             \nonumber \\
&&~~~~~~~~~+ \frac{1}{3}\int_{\Omega} f^{abc}
      (\Pi + \frac{\kappa}{8\pi} Q\cdot A)^a \phi^b Q\cdot\nabla \phi^c
 \nonumber \\
&&~~~~~~~~~+   \frac{\pi}{3\kappa} \int_{\Omega} f^{abc} f^{ars}
                 (\phi^b \Pi^c \phi^r \Pi^s) 
\nonumber \\
&&~~~~~~~~~+  \frac{\kappa}{576\pi}
\int_{\Omega} f^{abc} f^{ars}
(\phi^b Q\cdot\nabla \phi^c)(\phi^r Q\cdot\nabla \phi^s)\nonumber\\
&&~~~~~~~~~-{1\over 12}\int_{\Omega} f^{abc} f^{ars} (\phi^b \Pi^c \phi^r Q\cdot A^s
+ \phi^r Q\cdot A^s  \phi^b \Pi^c )+\cdots
\Biggr]
\label{ham2}
\end{eqnarray}
We must now consider the implementation
of the Gauss law.
This can also be done as a power series
in $f^{abc}$. Notice that the Gauss law operator
${\cal{G}}^a$ has the form
\begin{eqnarray}
{\cal{G}}^a &=& {\cal{G}}^a_0 +{\cal{G}}^a_1 +{\cal{G}}^a_2
                 + \cdots
\nonumber \\
{\cal{G}}^a_0 &=& \nabla \cdot E^a 
            -\frac{\kappa}{4\pi}\int_\Omega Q\cdot\nabla\phi^a
\nonumber \\
{\cal{G}}^a_1 &=& f^{abc}(A^b \cdot E^c +
                         \int_\Omega \phi^b \Pi^c)
\label{gauss2}
\end{eqnarray}
${\cal{G}}^a_0$ is of zero order in $f^{abc}$,
${\cal{G}}^a_1$ is of first order, etc.
We have also used (\ref{current}) to 
simplify (\ref{Gauss}).
Our strategy for the Gauss law
will be to find a unitary transformation
$S=e^F$ which has the property of transforming
${\cal{G}}^a$ to ${\cal{G}}^a_0$.
In other words we need
\begin{equation}
S^{-1} {\cal{G}}^a S \approx {\cal{G}}^a_0 
\label{unitary}
\end{equation}
where the weak equality means
``up to terms proportional to ${\cal{G}}^a_0$.'' Such a strategy
seems to have been known to
many people before; it has been used recently for Yang-Mills theories at 
zero temperature
in \cite{haller}; a similar approach at finite temperature but without
Debye mass was also used in \cite{landshoff} to clarify certain
technical questions regarding the imaginary-time formalism.
We first define a quantity
\begin{equation}
\Delta^a = \nabla\cdot A^a +
          2 \int_{\Omega} \frac{1}{(Q\cdot p)^2}
                 Q\cdot\nabla\Pi^a
\label{gf}
\end{equation}
$\Delta^a$may be considered as the gauge-fixing
constraint conjugate to ${\cal{G}}^a_0$.
Indeed we find
\begin{equation}
{[{\cal{G}}^a_0(x), \Delta^b (y) ]} = -i \delta^{ab}
      (-\nabla_x^2 + m_D^2) \delta(x-y)
\label{gfcomm}
\end{equation}
where $ m_D = \sqrt{2\kappa}$ is the Debye mass.
The inverse of the operator  on the right hand side
of (\ref{gfcomm}) will be denoted by
$G(x, y)$, i.e.,
\begin{equation}
G(x, y) = \int \frac{d^3 p}{(2\pi)^3}      
               \frac{e^{ip\cdot (x-y)}}{p^2 + m_D^2}
\label{green}
\end{equation}
Defining
\begin{equation}
\chi^a (x) = \int d^3 y ~G(y, x) \Delta^a (y)
\label{charge}
\end{equation}
we see that
\begin{equation}
{[{\cal{G}}^a_0(x), \chi^b (y)]} = -i \delta^{ab}\delta(x-y)
\end{equation}
Writing $ S = e^F \simeq 1 + F $,
to the first nontrivial order,
(\ref{unitary}) reduces to
\begin{equation}
{[{\cal{G}}^a_0(x), F ]} + {\cal{G}}^a_1 \approx 0
\end{equation}
The solution to this equation is given by
\begin{eqnarray}
F  &=& -i \int \chi^a f^{abc} (A^b \cdot E^c + \int_{\Omega}
                             \phi^b \Pi^c)     
-\frac{i}{2} \int \chi^a \partial_i \chi^b f^{abc} \tilde{E}_i^c
+ \cdots
\nonumber \\
\tilde{E}_i^c &=& E_i^c - \frac{\kappa}{4\pi}
                          \int_{\Omega} Q_i \phi^a
\label{f}
\end{eqnarray}
(Notice that $\nabla\cdot\tilde{E} = {\cal{G}}^a_0 $.)

The physical states can now be easily
constructed.
Let $|\psi_0 \rangle$ be any state obeying the condition
${\cal{G}}^a_0 |\psi_0 \rangle =0$.
Such states, as we shall see shortly, can be
constructed by plasmon creation operators
acting on the vacuum.
Equation (\ref{unitary}) then shows that physical states
$|\psi \rangle$ may be written as 
\begin{equation}
|\psi \rangle = S |\psi_0 \rangle
\end{equation}    
In terms of such physical states  we have for
the matrix element of ${\cal H}$
\begin{equation}
\langle\psi_1|{\cal H}|\psi_2 \rangle=\langle\psi_{1(0)}|S^{-1}{\cal H}
 S | \psi_{2(0)} \rangle
\end{equation}    
so that, in terms of the perturbatively
constructed states  $|\psi_0 \rangle$, the Hamiltonian
is effectively given by
\begin{equation}
{\cal H}_{eff} = S^{-1} {\cal H} S
\end{equation}
${\cal{G}}^a_0$ and $\Delta^a$ or ($\chi^a$)
will commute with the creation-annihilation operators
for the plasmons and, after obtaining ${\cal H}_{eff}$,
we can set ${\cal{G}}^a_0 =0$, $\Delta^a =0$.
Using the expression (\ref{ham2}) for the
Hamiltonian and $F$ as given 
by equation  (\ref{f}) we find
\begin{equation}
{\cal H}_{eff} = {\cal H}_0 + {\cal H}_{int} 
\end{equation}
with
\begin{eqnarray}
{\cal H}_0 &=&  \int \frac {E^2 + (\epsilon_{ijk}\partial_j A_k)^2}{2}
\nonumber \\
 & & +\frac{4\pi}{\kappa}\int_{\Omega}
                (\Pi - \frac{\kappa}{4\pi}Q\cdot A)^2
 + \frac{\kappa}{16\pi}\int_{\Omega}(Q\cdot\nabla \phi)^2
\end{eqnarray}
and
\begin{eqnarray}
{\cal H}_{int} &=&   \int f^{abc} \partial_i A_j^a A_i^b A_j^c  
      +\frac{1}{4} \int f^{abc}  f^{ars}
                        A_i^b A_j^c A_i^r A_j^s
\nonumber \\
                & & + \frac{1}{3} \int f^{abc}
                             (\Pi +\frac{\kappa}{8\pi}Q\cdot A)^a
                             \phi^b Q\cdot\nabla\phi^c
\nonumber \\
 & & +  \frac{2}{3}\int f^{abc}(\Pi + \frac{\kappa}{8\pi}Q\cdot A)^a
             \Bigl[ \phi^b G(x,y) \tilde{\rho}^c (y) 
\nonumber \\
     & &   -  \frac{1}{(Q\cdot p)^2} Q\cdot\nabla_x G(x,y)
            \tilde{\rho}^b (y) Q\cdot\nabla_x \phi^c (x) \Bigr]
\nonumber \\
            & & + \frac{\pi}{3\kappa}
                  \int f^{abc} f^{ars} \phi^b \Pi^c \phi^r \Pi^s
\nonumber \\
              & & + \frac{\kappa}{576\pi}
          \int f^{abc} f^{ars}
           (\phi^b Q\cdot\nabla \phi^c)(\phi^r Q\cdot\nabla \phi^s)
\nonumber \\
            & & -{1\over 12}\int_{\Omega} f^{abc} f^{ars} (\phi^b \Pi^c \phi^r Q\cdot
A^s + \phi^r Q\cdot A^s  \phi^b \Pi^c )\nonumber\\
 &&+\frac{1}{2} 
               \int \tilde{\rho}^a (x) G(x, y)
                    \tilde{\rho}^a (y) + \cdots
\label{Heff}
\end{eqnarray}
where
\begin{equation}
\tilde{\rho}^a = f^{abc} (A^b \cdot E^c +
                         \int_\Omega \phi^b \Pi^c)
\end{equation}
The interaction part of the Hamiltonian
now contains the screened Coulomb interaction
between plasmons.
 
\section{ Plasmon operators}

We now return to the construction of the plasmon 
operators which are eigenstates of the quadratic part
of the Hamiltonian, viz., ${\cal H}_0$.
Such operators must evidently obey the equation
\begin{equation}
{[ {\cal H}_0 , \alpha_{\lambda}^{\dagger a} (p) ]} =
\omega_{\lambda}(p) \,  \alpha_{\lambda}^{\dagger a} (p)
\label{cr}
\end{equation}
so that many-plasmon states are obtained
by multiple applications of $\alpha_{\lambda}^{\dagger a} (p)$
on the vacuum obeying the condition
$\alpha_{\lambda}^{a} (p) |0\rangle =0$.
In (\ref{cr}), $ \lambda $ is the polarization index.
For a given spatial momentum $\vec{p}$, we define
a triad of unit vectors $e_i^{\lambda}$, $\lambda = 1, 2, 3$, by
\begin{eqnarray}
e^3_i = \frac{p_i}{\sqrt{p^2}} & & 
          e^{\lambda}\cdot e^{\lambda'} = \delta^{\lambda \lambda'}
\nonumber \\
e^{\lambda}_{i} \, e^{\lambda'}_{j} \, \epsilon_{ijk}
& = & \epsilon^{\lambda \lambda' \alpha} e^{\alpha}_k
\end{eqnarray}
One can, if desired, make an explicit choice
\begin{equation}
e^1_i = \left( \frac{p_2}{\sqrt{p_1^2 +p_2^2}},
               \frac{-p_1}{\sqrt{p_1^2 +p_2^2}}, 0 \right),\quad
e^2_i = \frac{\left( p_3\, p_1, p_3\, p_2, -(p_1^2 +p_2^2) \right)}
          {\sqrt{(p_1^2 +p_2^2) \, p^2} }
\end{equation}
The plasmon creation operators may be taken 
to be of the form
\begin{equation}
\alpha_{\lambda}^{\dagger a} (p) =
\int d^3 x ~e^{-i p \cdot x}
\left[ e^{\lambda} \cdot ( \omega A^a - i E^a) +
\int_{\Omega}(f_1 \Pi^a + f_2 \phi^a) \right]
\label{plasmon}
\end{equation}
Substituting this in (\ref{cr}) and simplifying,
we see that solutions exist if 
$\omega_{\lambda}(p)$ are specific functions
of $\vec{p}$ obeying certain dispersion relations.
For the longitudinal plasmons, corresponding to
$e_i^3$, we find that $\omega_L = \omega_3$ 
is given by
\begin{equation}
1= \frac{\kappa}{2\pi} \int d\Omega
\frac{(Q\cdot e^3)^2}{\omega_L^2 -({Q}\cdot {p})^2}
\end{equation}
For the transverse polarizations we find
\begin{equation}
\omega_T^2 - p^2 = \omega_T^2 \,\frac{\kappa}{2\pi} 
\int d\Omega
\frac{(Q\cdot e^{\lambda})^2}{\omega_T^2 -({Q}\cdot {p})^2}
\end{equation}
The creation and annihilation operators, 
with the appropriate solutions for $f_1$, $f_2$
substituted into (\ref{plasmon}) are, for $\lambda =1, 2, 3$,
\begin{eqnarray}
\alpha_{\lambda}^{\dagger a} (p) &=&
\int d^3 x \frac{N_{\lambda}(p)  e^{-i p \cdot x}}
                {\sqrt{2\omega_{\lambda} V}}
\left[ e^{\lambda} \cdot ( \omega A^a - i E^a) +
\int d \Omega ~ f_{\lambda}  
(-\omega \Pi^a -\frac{i\kappa}{8\pi}({Q}\cdot {p})^2
  \phi^a) \right]
\nonumber \\
\alpha_{\lambda}^{a} (p) &=&
\int d^3 x \frac{N_{\lambda}(p)  e^{i p \cdot x}}
{\sqrt{2\omega_{\lambda} V}}
\left[ e^{\lambda} \cdot ( \omega A^a + i E^a) +
\int d \Omega ~f_{\lambda}  
(-\omega \Pi^a +\frac{i\kappa}{8\pi}({Q}\cdot {p})^2
  \phi^a) \right]
\label{ops}
\end{eqnarray}
where
\begin{eqnarray}
f_{\lambda} &=&
\frac{2 ~Q\cdot e^{\lambda}}
 {\omega_{\lambda}^2 -( {Q}\cdot {p})^2}
\nonumber \\
N_{\lambda}(p) &=&
\left[ 1+ \frac{\kappa}{8\pi}
     \int d \Omega (Q\cdot p)^2
f_{\lambda}^2 \right]^{-1/2}
\label{def2}
\end{eqnarray}
We have used plane-wave normalization
appropriate to a cubical box of volume
$V=L^3$, so that
$\vec{p}= (2\pi /L)\vec{n}$, 
$n_i \in {\bf Z}$.
The operators $\alpha, \alpha^\dagger$ obey the expected commutator algebra
\begin{equation}\label{CR}
{[\alpha_{\lambda}^{a} (p),
 \alpha_{\lambda'}^{b \dagger} (p') ]} = \delta^{ab}
\delta_{\lambda \lambda'}
\delta_{\vec{p}, \vec{p}'}
\end{equation}
with all other commutators equal to zero.
These operators create physical excitations
in the sense that
\begin{equation}
{[ {\cal{G}}_0^a (x), \alpha_{\lambda}^{b \dagger} (p) ]}
=0
\end{equation}
The operators $S^{-1} \alpha_{\lambda}^{a \dagger} (p) S$
are then operators which commute with Gauss law.
We can also verify that
$\Delta^a (x)$ commutes with
$\alpha$, $\alpha^{\dagger}$.
Thus for matrix elements with states built up of $\alpha^\dagger$'s, it is
consistent to set ${\cal G}_0^a =0,~\Delta^a =0$.

We now turn to the question of expressing the Hamiltonian (\ref{Heff})
in terms of the plasmon operators. The plasmons are collective mode excitations
in the plasma. In general, one can write every operator, such as $A,E,\phi, \Pi$,
in terms of $\alpha, ~\alpha^\dagger$ plus operators $\beta, ~\beta^\dagger$ which represent
other type of excitations.
If the nonplasmon excitations are weakly coupled to $\alpha, \alpha^\dagger$, then
an effective theory for the plasmons can be obtained by keeping just the 
$\alpha, \alpha^\dagger$-terms in $A,E,\phi ,\Pi$. This is, in general, how we could obtain an
effective theory for any kind of collective excitations. However, if
$\alpha,
\alpha^\dagger$ form a complete set of operators, clearly this would not be an approximation,
since there would be no independent $\beta, \beta^\dagger$-type modes anyway. As far as
$A_i,~E_i$ are concerned, it is easily seen that the plasmons do form a complete set. The
operators
$\phi (x,{\vec Q}),~\Pi (x,{\vec Q})$ have, in principle, an infinity of fields
defined just on spacetime; these may be viewed as the coefficients of the
expansion of $\phi,
\Pi$ in terms of spherical harmonics constructed from ${\vec Q}$. 
Thus completeness in terms of
$\alpha, ~\alpha^\dagger$ may seem doubtful. However, most of the degrees of freedom
corresponding to the arbitrariness of the ${\vec Q}$-dependence are irrelevant to us. This is
easily seen by going back to the action (\ref{WZW}) and eliminating $G$ by its equation of
motion. As shown elsewhere, there are no new propagating degrees of freedom in $G$. In fact, 
the elimination of $G$ leads to a purely $A$-dependent action and the analysis of this
action shows that the only physical modes are the plasmons. (The action, with $G$ eliminated,
is nonlocal in time making a Hamiltonian analysis difficult; this is why we need the auxiliary
field.)
We may therefore use the plasmons as the only modes relevant to the interactions and reduce 
${\cal H}$ entirely in terms of $\alpha, \alpha^\dagger$. First of all, we write $A_i^a(x)$
as
\begin{equation}
A_i^a(x)= \sum_{\lambda ,p}\left[ a_{\lambda i} \alpha_{\lambda}^a (p) +
a^*_{\lambda i} \alpha_{\lambda}^{a \dagger} (p) \right]
\label{A}
\end{equation}
Then, from (\ref{CR}), we find $a_{\lambda}(p)= -[\alpha_{\lambda}^{a\dagger},
A_i^a(x)]$ and using (\ref{ops}),
\begin{equation}
a_{\lambda i}(p)= {N_{\lambda }e^{-ip\cdot x}\over \sqrt{2\omega_{\lambda}V}}~e_i^{\lambda}
\end{equation}
We thus have
\begin{equation}
A_i^a (x)= \sum_{\lambda ,p}{N_{\lambda }(p) \over \sqrt{2\omega_{\lambda}V}}
\left[ \alpha_{\lambda}^a (p) e_i^{\lambda}e^{-ip\cdot x}
+\alpha_{\lambda}^{a\dagger}(p) e_i^{\lambda} e^{ip\cdot x}\right]
\label{mod1}
\end{equation}
In a similar way we find
\begin{eqnarray}
E_i^a(x)&=&\, \sum_{\lambda ,p}{N_{\lambda }(p) \over \sqrt{2\omega_{\lambda}V}}
(-i\omega_{\lambda})\left[ \alpha_{\lambda}^a (p) e_i^{\lambda}e^{-ip\cdot x}
-\alpha_{\lambda}^{a\dagger}(p) e_i^{\lambda} e^{ip\cdot x}\right]\\
\label{mod2}
\phi^a(x)&=&\, \int d^3y~ K_i(x,y)E_i^a(y)\\
\label{mod3}
\Pi^a&=&\, {\kappa \over 8\pi} \int d^3y~ ({\vec Q}\cdot \nabla )^2 K_i(x,y) A_i^a(y)
\label{mod4}
\end{eqnarray}
where
\begin{equation}
K_i (x,y,{\vec Q})= \int {d^3p\over (2\pi )^3} e^{ip\cdot (x-y)} \sum_{\lambda} 
{2{\vec Q}\cdot e^{\lambda} e_i^{\lambda} \over {\omega_{\lambda}^2 -
({\vec Q}\cdot {\vec p})^2}}
\end{equation}
Expressions (\ref{mod1}- \ref{mod4}) are easily checked to be compatible with the conditions
${\cal G}_0^a=0,~\Delta^a=0$. The Hamiltonian (\ref{Heff}) can now be expanded in terms of 
$\alpha,~\alpha^\dagger$ in a completely straightforward way; we just have to substitute
the mode expansions (\ref{mod1}-\ref{mod4}) in (\ref{Heff}).

\section{Discussion}

We shall conclude with some comments on the treatment of the auxiliary variables
$\phi ,\Pi$. The conditions $\Pi (x,-{\vec Q})=-\Pi (x,{\vec Q})$ and 
$\phi (x,-{\vec Q})=- \phi (x,{\vec Q})$ have to be imposed as weak conditions.
Otherwise one may get inconsistencies in the application of the current commutation
rules. Consider the expansion of the fields into harmonics on the two-sphere
given by ${\vec Q}$. We may write
\begin{eqnarray}
\phi (x, {\vec Q})&&= \sum_{lm} \phi_{lm} (x) ~Y_{lm}({\vec Q})\nonumber\\
\Pi (x, {\vec Q})&&= \sum_{lm} \Pi_{lm} (x) ~Y_{lm}({\vec Q})
\label{d1}
\end{eqnarray}
The canonical commutation rules show that $[\phi_{lm} (x), \Pi_{l'm'}(x')]
=i\delta_{ll'}\delta_{mm'} \delta (x-x')$. The requirement of these fields being odd
under ${\vec Q}\rightarrow -{\vec Q}$ is equivalent to
$\Pi_{lm}\approx 0,~\phi_{lm}\approx 0$ for even values of $l$.
$\Pi_{lm}\approx 0$ for even $l$ are first class constraints in the parlance of
constraint analysis
and the conditions $\phi_{lm}\approx 0$ may be taken as the gauge-fixing constraints for them.
One may set them to zero strongly one we have redefined commutators via Dirac brackets.
Since the other fields $\chi = (A_i,E_i, \phi_{lm}, \Pi_{lm})$ (for odd $l$)
commute with the constraints, the Dirac brackets are the same as the Poisson  brackets.
Thus we may set the constraints to zero at this stage.

In the Hamiltonian, the quadratic terms have no mixing of the even $l$- and odd $l$-components
of the fields. In the interaction terms, there is mixing but
we see that commutators of
$\chi$'s  always generate terms involving the constraints and hence it is consistent to set the
constraints to zero in the expression for the Hamiltonian.

In summary, we have expressed the Hamiltonian with the HTL effects added in terms of plasmon
creation and annihilation operators. One can use this for calculations of
different types of
higher order corrections; one can also use this as a starting point for variational
calculations involving the  thermal distribution of plasmons. Some of these are
currently under way. 

\medskip
\leftline{\large \bf Acknowledgments}
\smallskip
\noindent 
VPN thanks E. Moreno for discussions and for pointing out a slight error in the
previously given expression for the Hamiltonian (\ref{ham2}).
This work was supported in part by the National Science
Foundation Grant Number PHY-9605216.


\begin{thebibliography}{99}

\bibitem{BP1} R. Pisarski, {\it Physica} A 158, 246 (1989);
{\it Phys. Rev. Lett.}
63, 1129 (1989); E. Braaten and R. Pisarski, {\it Phys. Rev.}
D 42, 2156 (1990); {\it Nucl. Phys.} B 337, 569 (1990);
{\it ibid.} B 339,
310 (1990); {\it Phys. Rev.}  D 45, 1827 (1992).

\bibitem{Taylor} J. Frenkel and J.C. Taylor, {\it Nucl. Phys.} B 334, 199 (1990);
J.C. Taylor and S.M.H. Wong, {\it Nucl. Phys.} B 346, 115 (1990).

\bibitem{VPN1} R. Efraty and V.P. Nair, {\it Phys. Rev. Lett.} 68, 2891 (1992);
{\it Phys. Rev.} D 47, 5601 (1993); R. Jackiw and V.P. Nair, {\it Phys. Rev.} 
D 48, 4991 (1993).

\bibitem{blaizot} J.P. Blaizot and E. Iancu, {\it Phys.Rev.Lett.} 70, 3376
(1993); {\it Nucl.Phys.}  B 417, 608 (1994).

\bibitem{manuel1} R. Jackiw, Q. Liu and C. Lucchesi, {\it Phys. Rev.}
 D 49, 6787 (1994); P.F. Kelly, Q. Liu, C. Lucchesi and C. Manuel,
{\it Phys. Rev. Lett.} 72, 3461 (1994); {\it Phys. Rev.}
D 50, 4209 (1994).
 
\bibitem{tft98} see, for example, various articles in {\it TFT98: Thermal Field Theory
and Applications}, U. Heinz (ed.), hep-ph/9811469.

\bibitem{braat} E. Braaten and A. Nieto, {\it Phys. Rev.} D 51, 6990 (1995);
J.O. Anderson, E. Braaten and M. Strickland, hep-ph/9902327, hep-ph/9905337.

\bibitem{GPY} A.D. Linde, Phys. Lett. B 96, 289 (1980);
D. Gross, R. Pisarski and L. Yaffe, Rev. Mod. Phys. 53, 43 (1981).

\bibitem{iancu} J.P. Blaizot and E. Iancu, hep-ph/9903389.

\bibitem{bodeker} D. B\"odeker, {\it Phys. Lett.}  B 426, 351 (1998);
hep-ph/9903478.

\bibitem{manuel2} P. Arnold, D. Son and L.G. Yaffe, hep-ph/9810216, hep-ph/9901304;
D. Litim and C. Manuel, hep-ph/9903462.

\bibitem{VPN2} V.P. Nair, {\it Phys. Rev.} D 48, 3432 (1993);
{\it ibid.} D 50, 4201 (1994).

\bibitem{haller} M. Belloni. L. Chen and K. Haller, {\it Phys. Lett.} B 403, 316
(1997). 

\bibitem{landshoff} K.A. James and P.V. Landshoff, {\it Phys. Lett.}
B 257, 167 (1990).

\end{thebibliography}
\end{document}